\documentclass[aps,prb,twocolumn,superscriptaddress,showpacs]{revtex4-2}
\usepackage{amsmath,amssymb,amsfonts,dsfont, float,graphics,epsfig,epstopdf,color,verbatim,tabularx,bm,multirow}
\usepackage{wasysym}
\usepackage[utf8]{inputenc}
\usepackage[T1]{fontenc}
\usepackage{xcolor}

\usepackage{hyperref}
\hypersetup{
	colorlinks = true,
	linkcolor = [rgb]{0.70,0.13,0.13},
	citecolor = [rgb]{0.13,0.55,0.13},
	urlcolor = [rgb]{0.25,0.41,0.88}}

\DeclareSymbolFont{sfletters}{OML}{cmbrm}{m}{it}
\DeclareMathSymbol{\sfeps}{\mathord}{sfletters}{"22}
\newcommand{\Tr}{\operatorname{Tr}}

\graphicspath{{Figures/}}

\begin{document}
\title{Scaling of Disorder Operator and Entanglement Entropy at Easy-Plane Deconfined Quantum Criticalities}

\author{Jiarui Zhao}
\email{jrzhao@connect.hku.hk}
\affiliation{Department of Physics and HKU-UCAS Joint Institute of Theoretical and Computational Physics, The University of Hong Kong, Pokfulam Road, Hong Kong SAR, China}
\affiliation{Department of Physics, The Chinese University of Hong Kong, Sha Tin, Hong Kong SAR, China}

\author{Zi Yang Meng}
\email{zymeng@hku.hk}
\affiliation{Department of Physics and HKU-UCAS Joint Institute of Theoretical and Computational Physics, The University of Hong Kong, Pokfulam Road, Hong Kong SAR, China}

\author{Yan-Cheng Wang}
\email{ycwangphys@buaa.edu.cn}
\affiliation{Hangzhou International Innovation Institute, Beihang University, Hangzhou 311115, China}
\affiliation{Tianmushan Laboratory, Hangzhou 311115, China}

\author{Nvsen Ma}
\email{nvsenma@buaa.edu.cn}
\affiliation{School of Physics, Beihang University, Beijing 100191, China}

\begin{abstract}
We systematically investigate the scaling behaviors of the disorder operator and the entanglement entropy (EE) of the easy-plane JQ (EPJQ) model at its transitions between the antiferromagnetic XY ordered phase (AFXY) and the valence bond solid (VBS) phase. We find there exists a tiny yet finite value of the order parameters at the AFXY-VBS phase transition points of the EPJQ model, and the finite order parameter is strengthened as anisotropy $\Delta$ varies from the Heisenberg limit ($\Delta=1$) to the easy-plane limit ($\Delta=0$). This observation provides evidence that the N\'eel-VBS transition in the JQ model setting evolves from weak to prominent first-order transition as the system becomes anisotropic. Furthermore, both EE and disorder operator with smooth boundary cut exhibit anomalous scaling behavior at the transition points, resembling the scaling inside the Goldstone mode (AFXY) phase, and the anomalous scaling becomes strengthened as the transition becomes more first order. In particular, for $\Delta \le 0.3$, the obtained log-coefficients converge to 0.5 which is the same as the contribution from one Goldstone mode in the N\'eel phase. For $\Delta > 0.3$, the log-coefficients are smaller and our findings might suffer from strong finite-size effects due to the fact that the remaining N\'eel order here is quite tiny.

\end{abstract}
\date{\today}
\maketitle

\section{Introduction}
The conventional Landau-Ginzburg-Wilson (LGW) paradigm for characterizing phases and their transitions has been confronted with the notion of deconfined quantum critical points (DQCPs), which describes a direct continuous transition between  N\'eel ordered phases and valence-bond-solid (VBS) phases~\cite{Senthil_2004,Senthil2004-2,levin2004deconfined,Senthil2005}. However, the feasibility of realizing such transitions in lattice models remains a contentious issue because of the observations of various anomalous behavior against conformal field theories (CFT)~\cite{harada2013possibility,chenDeconfined2013,nahum2015deconfined,nakayamaNecessary2016,Li:2018lyb,polandConformal2019,wangScaling2022,zhaoScaling2022,songExtracting2023,song2023deconfined,liuFermion2023,liaoTeaching2023,liuDisorder2024}, such as drifting of critical exponents~\cite{harada2013possibility,chenDeconfined2013,nahum2015deconfined}, violation of CFT bounds~\cite{nakayamaNecessary2016,Li:2018lyb,polandConformal2019}, and the anomalous scaling behavior of entanglement entropies and disorder operators~\cite{wangScaling2022,zhaoScaling2022,songExtracting2023,song2023deconfined,liuFermion2023,liaoTeaching2023,liuDisorder2024}. A deeper exploration of the origins of these anomalous behaviors will benefit not only the better understanding of nature of DQCPs but also its lattice model and even material realizations~\cite{zayedSpin2017,guoQuantum2020,jimenezQuantum2021,sunEmergent2021,cuiProximate2023,guoDeconfined2023}.

Among all the anomalous behavior against CFTs,  we are in particular interested in the anomalous finite-size scaling behavior of R\'enyi entanglement entropy (EE) and disorder operator, as it happens even at relatively small system sizes~\cite{wangScaling2022,zhaoScaling2022,songExtracting2023,song2023deconfined,liuFermion2023,liaoTeaching2023,liuDisorder2024} where other physical quantities such as order parameters and correlation functions still exhibit good agreement with a continuous phase transition~\cite{Sandvik2007,louAntiferromagnetic2009,shaoQuantum2016,maDynamical2018,qinDuality2017}. In fact, at the DQCPs of SU(2) JQ$_2$ and JQ$_3$ models, our previous works~\cite{zhaoScaling2022,songExtracting2023,wangScaling2022,song2023deconfined} show that disorder operator and EE all exhibit anomalous scaling against CFTs. To be specific, for the corner cut case where the boundary of subregion has four $\frac{\pi}{2}$ corners, both disorder operator and EE have a positive logarithmic correction to the leading order area law term which violates the unitary CFT prediction that the correction must be negative ~\cite{casiniPositivity2012,FradkinPRL2006,Casini_2007}. Later, further studies clarified that the violation of EE actually arises from the smooth part of the subregion. In fact, for smooth cut case where the boundary has no sharp corners, the anomalous logarithmic subleading correction also exists in EE, which is incompatible with CFTs~\cite{song2023deconfined,songExtracting2023}, but seems to be more compatible with the existence of Goldstone modes ~\cite{metlitskiEntanglement2011,dengDiagnosing2024} and can also possibly be adapted with the `walking" pseudo-criticality behavior at the transition~\cite{wang2017deconfined,NahumPRB2020,RCMa2020,zhou2023mathrm}. In addition, recently another work~\cite{demidioEntanglement2024} shows that scaling of EE at the DQCP is cut-dependent, and the anomalous log correction for smooth cut can be suppressed by considering a tilted square lattice by $\frac{\pi}{4}$ degrees, and the essential information of the CFTs might be captured by this cut at small system sizes. In this context, the scalings of these nonlocal physical observables  prove to be powerful and sensitive tools for diagnosing the various possible scenarios of the DQCPs~\cite{zhaoScaling2022,songExtracting2023,wangScaling2022,song2023deconfined, dengDiagnosing2024} and revealing the universal information of the possible CFT nearby the transitions~\cite{demidioEntanglement2024}.

Here in this work, we aim to shed more insights on this issue by tunning the JQ-type N\'eel-VBS transition from weakly first-order~\cite{takahashiSO52024} to a more prominent first-order transition, such that many of the salient features observed at the SU(2) JQ limit, for example, the anomalous finite-size scaling behavior in the entanglement entropy (EE)~\cite{zhaoScaling2022,songExtracting2023,song2023deconfined,dengDiagnosing2024} and disorder operator~\cite{wangScaling2022}, can also be examined at a clear first-order transition, and the connection between the observed anomalous behavior in EE and disorder operator with the first-order nature of the transitions can be further clarified.
A possible platform to achieve this goal is the easy-plane JQ (EPJQ) model, which adds the easy-plane anisotropy to the Heisenberg term in the JQ model, as shown in Eq.~\eqref{eq:eq1}, to tune the N\'eel-VBS transition to be more first-ordered. Previous studies have investigated the nature of the N\'eel-VBS transition at different anisotropy $\Delta$, both in 2D and 3D lattice model settings~\cite{qinDuality2017,maRole2019,zhaoSymmetry2019,desaiFirst2020,sunEmergent2021}, and the general expectation/observation is that the more $\Delta$ tunes away from the Heisenberg limit ($\Delta=1$) towards the easy-plane limit ($\Delta=0$), the stronger the first order of the N\'eel-VBS transition. The EPJQ model thus serves as an ideal platform to study the anomalous scaling behavior of EE and disorder operator at these transition points simply by tuning the values of $\Delta$. 

In this paper, we use the stochastic series expansion (SSE) quantum Monte Carlo (QMC) methods~\cite{Sandvik1999,Syljuaasen2002} to systematically study the order parameter, entanglement entropy, and disorder operator at the transitions of the EPJQ model. We carefully analyze the remaining ordered moments at the transitions and systematically study the evolution 
 of the scaling of entanglement entropy and disorder operator along the critical line as a function of $\Delta$ in Eq.~\eqref{eq:eq1}. Our major findings are

 \begin{enumerate}
     \item For $\Delta \in [0,1)$ there always exist tiny yet finite  order parameters at the AFMXY/N\'eel-VBS phase transitions of the model (including the finite order parameter at the $\Delta=1$ limit~\cite{takahashiSO52024}), indicating the first-order nature of the transition, and the moment increases as $\Delta$ is tuned from 1 to 0;
     \item Both EE and disorder operator with standard smooth boundary cut (without tilting) exhibit anomalous scaling behavior against CFTs at the transition points, which resemble the the scalings inside the Goldstone model phase, and the anomalous scalings also become strengthened as the transition becomes more first order.
     \item For $\Delta \le 0.3$, the obtained log-coefficients converge to 0.5 which is the same as the contribution from one Goldstone mode in the N\'eel phase. For $\Delta > 0.3$, the log-coefficients are smaller than 0.5, and that might be caused by strong finite-size effects.
 \end{enumerate}


Our work thus provides a comprehensive study of the scaling of EE and disorder operator along the critical line where the weakly first-order transition is gradually tuned to a prominent first-order one by tunning $\Delta=1$ to $\Delta=0$. We show that the anomalous scalings of EE and disorder operator are very positively related with the first order behavior,suggesting that the DQCP of JQ model is indeed a very weakly-first-order transition and the anomalous log-corrections come from Goldstone modes.

The rest of the paper is organized as follows. We first introduce the EPJQ model and discuss how we extract the finite order parameters at its transition points, then we move on to the discussion of disorder operator and reveal its anomalous scaling with comparisons to that of the conventional (2+1)D O(3) QCP and N\'eel phase, next we discuss the results of EE at the EPJQ transitions  and point out its anomalous scaling very likely stems from the residual Goldstone mode at the first-order transition points, finally we give a comprehensive summary on our results and point out future directions. Detailed analysis of the determination of the transition points and extrapolations of the order parameter, and the analysis of the quality of the fitting in EE data are presented in the Appendix ~\ref{sec:SMII}.\\

\begin{figure*}[htp!]
\centering
\includegraphics[width=\textwidth]{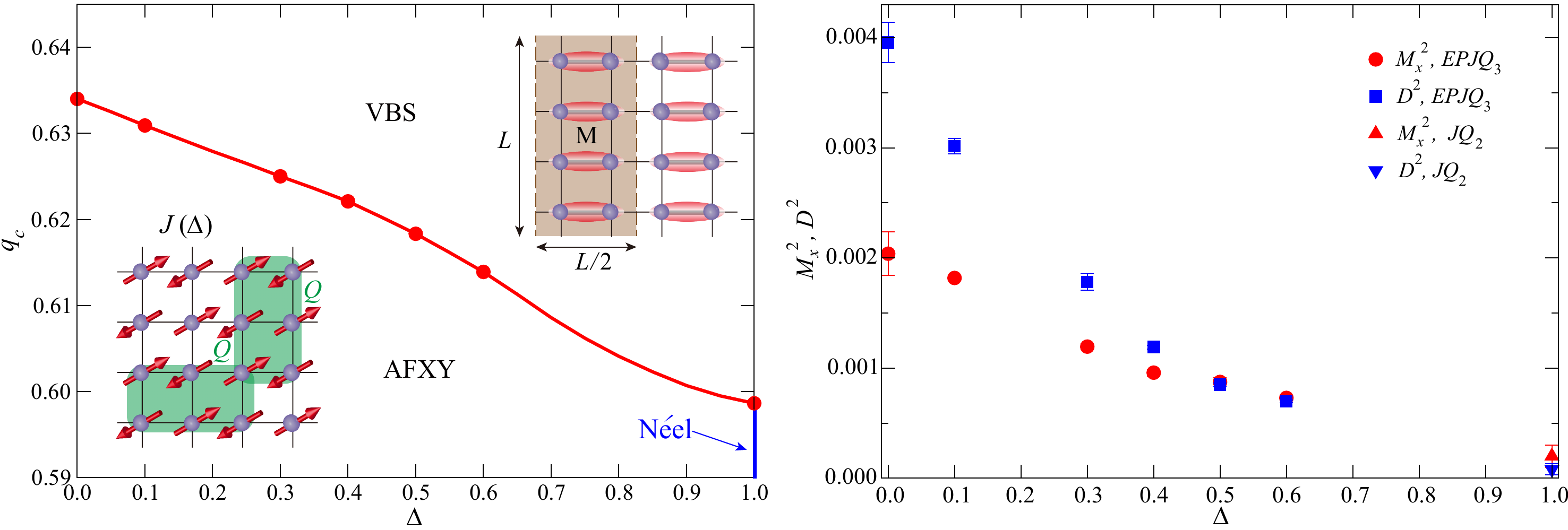}
\caption{Phase diagram of the EPJQ model and finite order parameter at the transitions. (a) Ground state phase diagram of the easy-plane JQ$_3$ model. The red dots are critical points $q_c$ determined from FSS analysis of Binder cumulants for different $\Delta$ values. When $\Delta=1$, the model is the standard JQ$_3$ model and the weakly-first-order transition point~\cite{takahashiSO52024} separates the N\'eel order state with the VBS state. At finite $\Delta$, the antiferromagnetic N\'eel order is of the XY type. The N\'eel-VBS transition becomes prominently first order as $\Delta$ goes from 1 to 0. The insets exhibit the EPJQ model (left) and the entanglement region with smooth boundaries (right). (b) The non-zero order parameters for both AFXY and VBS phases at $q_c(\infty)$ with $\Delta$ changes from $0$ to $0.6$. At $\Delta=1$, the extrapolated order parameters are from Ref.~\cite{takahashiSO52024} of JQ$_2$ model.}
\label{fig:fig1}
\end{figure*}

\section{Model and Phase Diagram.}
\noindent We study the easy-plane JQ$_3$ (EPJQ) model~\cite{qinDuality2017,maRole2019,zhaoSymmetry2019,desaiFirst2020,sunEmergent2021} as illustrated in the lower left inset of Fig.~\ref{fig:fig1} (a) with the following Hamiltonian
\begin{equation}
H = J\sum_{\langle ij \rangle} \left(S_i^xS_j^x+S_i^yS_j^y+\Delta S_i^zS_j^z \right) - Q\sum_{\langle ijklmn \rangle} P_{ij}P_{kl}P_{mn},
\label{eq:eq1}
\end{equation}
 where $P_{ij}=\frac{1}{4}-\mathbf{S}_{i}\cdot\mathbf{S}_{j}$ is the two-spin singlet projector. For $\Delta=1$, the model reduces to JQ$_3$  model with isotropic Heisenberg interactions. 
 It has been found a direct quantum phase transition in the JQ$_3$ model between the N\'eel and VBS phases happens at $q_{c}=0.59864(5)$ (with $q=\frac{Q}{J+Q}$ and $J+Q=1$ as the energy unit)~\cite{louAntiferromagnetic2009,wangScaling2022}. However,  more numerical evidence, especially from the scaling of nonlocal observables such as EE~\cite{zhaoScaling2022,songExtracting2023,song2023deconfined} and disorder operator~\cite{wangScaling2022}, reveal an anomalous behavior against CFTs even if the partitioning of the lattice is smooth, i.e. no sharp corners on the boundary. The fact that the sign of the observed log-coefficient is consistent with that of the presence of the Goldstone mode, further suggests there exist finite antiferromagnetic moment, i.e. remaining of the N\'eel order, at the transition point~\cite{metlitskiEntanglement2011}. Such evidence promotes the understanding that the N\'eel-VBS transition is indeed first order, despite being very weak at the SU(2) limit~\cite{takahashiSO52024}. In this work, we  monitor the behavior of both conventional  order parameters and nonlocal observables at such transitions, as the anisotropy $\Delta$ in Eq.~\eqref{eq:eq1} is tuned from the Heisenberg limit($\Delta=1$) to easy-plane limit ($\Delta=0$). 
 
\begin{figure*}[htp!]
\centering
\includegraphics[width=\linewidth]{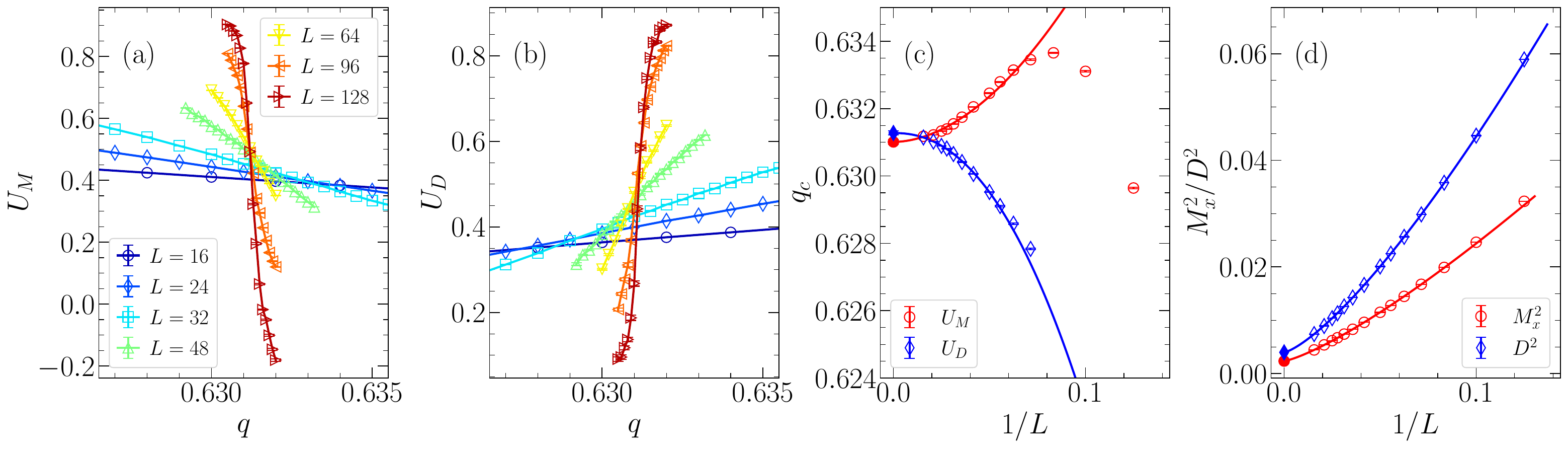}
\caption{ Determination of phase transition points and extrapolated order parameter at the transitions.
The size dependence of crossing points of Binder cumulant of (a) N\'eel and (b) VBS order parameter in easy-plane JQ$_3$ model for $\Delta=0.1$. The fact $U_M(L=128)$ becomes negative close to the transition suggests its first-order nature. (c) The critical points at the thermodynamic limit for $1/L=0$ are obtained through the crossing point analysis of all the $q_{c}(L)$ using Eq.~\eqref{eq:eq5}. (d)The squered N$\acute{e}$el and VBS order parameters as functions of $1/L$ at $\Delta=0.1$ cross-points with polynomial fits for both orders with the residual order at the thermodynamic limit. All the residual moments of EPJQ model in Fig.~\ref{fig:fig1} (b) are obtained in this way.}
\label{fig:fig2}
\end{figure*}

\begin{table*}[!t]
    \centering
    \begin{tabular}{c|c|c|c|c|c|c}
        \hline

             &$\Delta=0$    & 0.1& 0.3 &0.4 &0.5 & 0.6  \\
        \hline
        $q_{c}$ using $U_{M}$  & ~0.6340(1)~ & ~0.63091(5)~ & ~0.6250(2)~ & ~0.6221(2)~ & ~0.61883(6)~  & 0.6139(3)      \\
        \hline
        $q_{c}$ using $U_{D}$ & ~0.6341(5)~ & ~0.6311(9)~ & ~0.6261(4)~ & ~0.6225(2)~ &  ~0.61883(5)~ & ~0.6142(1)~   \\
        \hline
    \end{tabular}
    \caption{The critical points obtained from the Binder cumulants of $U_M$ and $U_D$ as shown in Fig.~\ref{fig:fig2} for different $\Delta=0,0.1, 0.3, 0.4, 0.5$ and 0.6 using similar analysis in Fig.~\ref{fig:fig2}. The $q_c(\infty)$ of $U_M$ and $U_D$ are identical within two sigma. The detailed analyses are shown in the Appendix~\ref{sec:SMI}.}
    \label{qc}
\end{table*}

 At each $\Delta \in [0,1)$, the phase transition between the antiferromagnetic XY (AFXY) and VBS phase at zero temperature happens when tunning $q=Q/(J+Q)$ from zero to $q_c$. To determine the critical points, we perform SSE-QMC simulations~\cite{Sandvik1999,Syljuaasen2002} on the EPJQ model at $\beta=2L$ for different system sizes $L=16,24,32,48, 64,96,128$ with $\Delta=0$, $0.1$, $0.3$, $0.5$ and $0.6$.  In the AFXY phase, the order parameter is defined as the sublattice magnetization which breaks the U(1) symmetry, with the  $x$ component being written as
 \begin{equation}
    \begin{split}
       M_{x}=\frac{1}{N}\sum_{i=1}^{N}(-1)^{x_{i}+y_{i}}S_{i}^{x},
        \label{eq:eq2}
   \end{split}
\end{equation}
where $\{x_{i}$,$y_{i}\}$ are coordinates of site $i$ in the lattice. In the simulation with finite system sizes, the expectation value of $\langle M_{x}^{2}\rangle$ can be viewed as the square of the AFXY order parameter, whereas $\langle M_x \rangle =0$ at finite sizes. 

For the VBS phase, the valence bonds can form in horizontal or vertical  directions as exemplified of the former in the right inset of Fig.~\ref{fig:fig1} (a). This order can be quantified through observables
\begin{equation}
\begin{split}
    D_{x}=\frac{1}{N}\sum_{i}(-1)^{x_i}S_{x_i,y_i}\cdot S_{x_i+1,y_i},\\
D_{y}=\frac{1}{N}\sum_{i}(-1)^{y_i}S_{x_i,y_i}\cdot S_{x_i,y_i+1},
\end{split}
\label{opall}
\end{equation}
with $\{x_i,y_i\}$ the coordinates of site $i$. In this way, the expectation value of the square of the order parameter in VBS phase is $\langle D^{2}\rangle=\langle D^{2}_{x}\rangle+\langle D^{2}_{y}\rangle$. 

We thus perform the finite-size scaling (FSS) analysis of $\langle M^2_x \rangle$ and $\langle D^2 \rangle$ to determine the transition points at the thermodynamic limit (TDL)~\cite{maAnomalous2018}. The dimensionless quantity Binder cumulants are widely used in the FSS to extract the critical points. For those two ground states in our study, the Binder cumulants are defined as 
\begin{equation}
\begin{split}
 U_{M}=2\left( 1-\frac{\langle M_{x}^{4}\rangle}{3\langle M_{x}^{2}\rangle^{2}}\right),\\
  U_{P}=2\left(1-\frac{\langle D^{4}\rangle}{2\langle D^{2}\rangle^{2}}\right),
  \end{split}
  \label{eq:eq4}
\end{equation}
where normalization factors and constants in Eq.~\eqref{eq:eq4} are decided from the degree of freedom and number of components for order parameters. Under this definition, $U_{M}\to 1$ in AFXY while it is zero in the VBS state when $L\to\infty$. On the contrary, in AFXY phase $U_{P}\to 0$ and $U_{P}\to 1$ in the VBS phase. Besides, both $U_{M}$ and  $U_{P}$ are dimensionless quantities whose value are independent of system sizes at critical points. In this way, for different simulated systems $U_{M}(q,L_{1})=U_{M}(q,L_{2})$ and $U_{D}(q,L_{1})=U_{D}(q,L_{2})$ at critical point $q_{c}$, which means that the $q$ dependence of Binder cumulants for different sizes should cross with each other at the critical point. 

Even though there only exist one phase transition point at thermodynamic limit (TDL) $q_{c}(\infty)$, usually all those curves will not exactly cross at  $q_{c}(\infty)$ because of the finite size effect and corrections. Therefore, we shall locate all the crossings $q_{c}(L)$ where $U_{M}(q,L)=U_{M}(q,2L)$ or $U_{D}(q,L)=U_{D}(q,2L)$ and trace the $q_{c}(\infty)$ using the following scaling form 
\begin{equation}
 q_{c}(L)=q_{c}(\infty)+aL^{-1/\nu-\omega}
  \label{eq:eq5}
\end{equation}
with $\nu$ the correlation length exponent and $\omega$ the finite-size correction exponent~\cite{qinDuality2017,maAnomalous2018,chenPhases2023}.

\begin{figure*}[t]
\centering
\includegraphics[width=\textwidth]{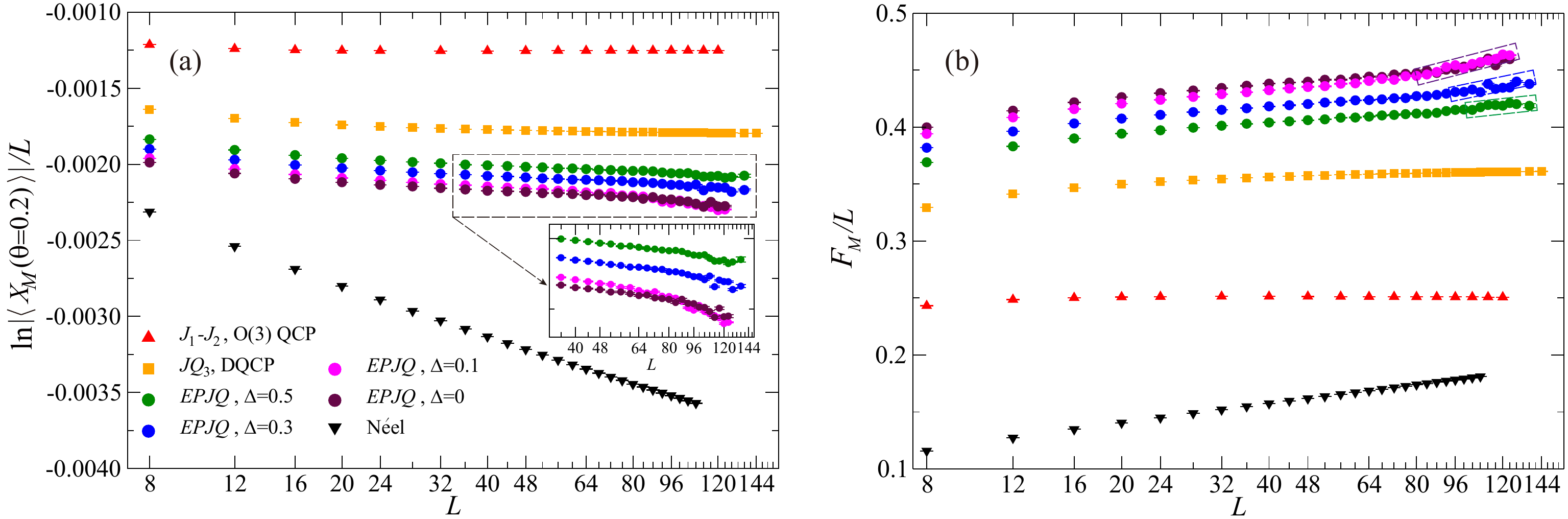}
\caption{Disorder operator bipartite spin fluctuations in various quantum states. (a) Disorder operator at $\theta=0.2$  and (b) the bipartite spin fluctuations $F_{M}$ as functions of system size $L$ for the N$\acute{e}$el phase, (2+1)D O(3) QCP of $J_1-J_2$ model (Eq.~\eqref{eq:eq6}), DQCP of JQ$_3$ and the first-order quantum phase transition points of EPJQ (Eq.~\eqref{eq:eq1}), respectively. The inset in (a) and the dashed rectangular boxes in (b) show the slopes of $\ln |X_M(\theta)|/L$ and $F_{M}$ became sharper as system size increases at the quantum phase transition points of EPJQ, signifying the first-order nature of the transition and such sharper slopes resemble that of the N\'eel phase with spontaneous symmetry breaking. (a) and (b) share the same data labeling for different models.}
\label{fig:fig3}
\end{figure*}

We perform such FSS analysis on the computed results in EPJQ model with different $\Delta$. For example, the Binder cumulants of both AFXY and VBS phases close to the crossing points are presented in Fig.~\ref{fig:fig2} (a) and (b) for $\Delta=0.1$. The obtained crossings of two different curves with $(L,2L)$ with $L$ increasing from $8$ to $64$ are then illustrated in Fig.~\ref{fig:fig2} (c) as well as the fitting results. Using the fitting form in Eq.~\eqref{eq:eq5} two $q_{c}(\infty)$ can be obtained with $q_{c}(L)$ related to different order parameters. In Fig.~\ref{fig:fig2} (d), we also present the extrapolation of the ordered moments (square) of $\langle M^2_x\rangle$ and $\langle D^2 \rangle$ at each $q_c(L)$ and eventually at the TDL, we find both $\langle M^2_x\rangle$ and $\langle D^2 \rangle$ are finite (despite small in the $y$-scale of the figure) at the transition point and this is the defining evidence that the transition is first-order.

Repeating such a procedure, we can also get $q_{c}(\infty)$ for different $\Delta$ in Tab.~\ref{qc}. For a given $\Delta<1$, the $q_{c}(\infty)$ obtained from crossings of different dimensionless quantities agree with each other considering the error as large as two sigma, which can be true for the continuous phase transition as well as the first-order one. However, we should point out that $U_{M}$ of the AFXY order parameter for $\Delta=0.1$ becomes negative close to $q_{c}$ at the largest simulated size $L=128$, as shown in Fig.~\ref{fig:fig2} (a), the negative $U_{M}$ can also be regarded as a signature of the first-order phase transition~\cite{binderFinite1984}. 

We have performed the extrapolation of the ordered moments as in Fig.~\ref{fig:fig2} (d) for all the $\Delta$ values, and the obtained results are summarized in Fig.~\ref{fig:fig1} (b). It is clear that $\langle M^2_x\rangle$ and $\langle D^2 \rangle$ are small at $\Delta=0.6$ (they are even smaller at the isotropic limit), but they gradually increases as $\Delta \to 0$. We show all these extrapolations in the Appendix~\ref{sec:SMI}. The enhancement of the ordered moments as $\Delta \to 0$ along the phase AFXY-VBS phase boundary in Fig.~\ref{fig:fig1} (a)  suggests evolution from  very weakly first-order to stronger ones. As will be shown below, our non-local EE and disorder operator measurements (with smaller systems sizes compared with  order parameters) can also capture such intriguing features. Note that the tiny remaining order parameters may appear to contradict the assumption of our FSS analysis that the system is critical and scale invariant. However, as suggested by our data and a reference by Takahashi et al.~\cite{takahashiSO52024} for larger system sizes, these weakly first-order transitions exhibit very long correlation lengths, which ensures that the assumption of scale invariance in our FSS analysis holds at the system sizes and precision we consider. It is worth noting that the scaling behavior of binder ratio at evident first order transitions has been discussed in Ref.~\cite{pelissetto2025}.


For comparison, we have also studied the scaling of the disorder operator for the square lattice $J_1$-$J_2$ Heisenberg model with the Hamiltonian
\begin{equation}
        H_{J_1-J_2}=J_1\sum_{\langle ij \rangle} \mathbf{S}_{i}\cdot\mathbf{S}_{j}+J_2\sum_{\langle ij \rangle^{'}} \mathbf{S}_{i}\cdot\mathbf{S}_{j},
        \label{eq:eq6}
\end{equation}
where $\langle ij \rangle$ denotes the $J_1$ bond and $\langle ij \rangle^{'}$ denotes the $J_2$ bond. This model has a well-established (2+1)d O(3) QCP at $(J_2/J_1)_c=1.90951(1)$~\cite{maAnomalous2018} separating the N\'eel phase and a symmetric singlet product phase. As will also shown below, the scaling of disorder operator of this O(3) QCP does not have anomalous log-corrections. \\

\section{Disorder Operator}
\noindent Once the transition points of each $\Delta$ are determined, we now carry out the analysis of the disorder operator upon them. As a non-local observable, disorder operator is defined as the expectation value of a symmetry transformation applied to a finite region in the statistical or quantum many-body systems of interest~\cite{Wegner1971,KadanoffPRB1971,fradkin2017disorder,nussinov2009sufficient,nussinov2009symmetry}. The design and implementation of the disorder operator and the analysis of its finite size scaling behavior have been successfully carried out in the situations of spontaneous symmetry breaking phase, quantum critical points, the symmetric phases with topological orders, symmetric mass generation transition and even the free fermion surface and interacting quantum critical Fermi surface systems~\cite{zhaoHigher2021,wuCategorical2021,wangScaling2021,xuUniversal2021,chenTopological2022,wangScaling2022,jiangMany2023,liuFermion2023,liuDisorder2024,caiDisorder2024,wuBipartite2024}.

In a 2D lattice spin model, for a region $M$ as shown in the right inset of Fig.~\ref{fig:fig1} (a), we define the U(1) disorder operator $X_M(\theta)=\prod_{i \in M}e^{i\theta (S^{z}_i+\frac{1}{2})}$, where $S_i^z$ is the U(1) charge on site $i$. For the  case of region M with sharp corners,  the scaling of the U(1) disorder operator have been studied systematically~\cite{wuCategorical2021,wangScaling2021,xuUniversal2021,jiangMany2023}.
In the ordered (U(1) or SU(2) symmetry breaking) phases, such as the superfluid phase or N$\acute{e}$el phase, it was found that $\ln |X_M(\theta)|\sim -bL\ln L$ ~\cite{wangScaling2021,wangScaling2022}. At the quantum critical points of 2D lattice models, previous studies ~\cite{zhaoHigher2021,wangScaling2021,wangScaling2022} showed that $\ln |X_M(\theta)|$ takes the following general form for a rectangle region $M$:
\begin{equation}
\label{eq:eq7}
    \ln |X_M(\theta)|=-a_1 L+s\ln(L)+a_0,
\end{equation}
where all the coefficients are functions of $\theta$ and the log-coefficient $s$ follows a universal function of both $\theta$ and the opening angles $\phi$ of the corners of $M$ ($\phi=\pi/2$ for rectangle region). Given a smooth region M (without corners on boundary), there should be no corner correction and the log-coefficient $s=0$. Such area-law decay of disorder operator, i.e. $\ln |X_M(\theta)|=-a_1 L+a_0$, holds both at the QCP and inside the gapped symmetric phases.

To detect the scaling of the disorder operator at the quantum phase transition point of EPJQ model in Eq.~\eqref{eq:eq1}, we choose the entanglement region $M$ to be a $L\times L/2$ cylinder region (without corners) in the lattice, as shown in the right inset of Fig.~\ref{fig:fig1} (a), with system size for $L=8$ up to $L=144$. For a good comparison between different quantum states, we calculated the U(1) disorder operator (with small rotation angle $\theta=0.2$) for the N\'eel phase (standard spin-1/2 Heisenberg model on the square lattice), the O(3) QCP of $J_1-J_2$ model in Eq.~\eqref{eq:eq6}, DQCP of JQ$_3$ model (Eq.~\eqref{eq:eq1} at $\Delta=1$), and quantum phase transition points of EPJQ of Eq.~\eqref{eq:eq1} for $\Delta=0, 0.1, 0.3, 0.5$, respectively. The results are shown in Fig.~\ref{fig:fig3} (a).

As the leading term of the U(1) disorder operator $X_M$ at small $\theta$ for the N$\acute{e}$el phase is $\ln |X_M(\theta)|\sim -bL\ln L$, one expects $\ln X_M /L$  proportional to $\ln L$ with a pronounced slope, as shown in Fig.~\ref{fig:fig3} (a). For the (2+1)D O(3) QCP of $J_1-J_2$ model, since $\ln |X_M(\theta)|\sim L$ with log-coefficient $s=0$ due to the smooth boundary,$\ln |X_M(\theta)|/L$ will be a constant at large system size, which is also clearly seen in Fig.~\ref{fig:fig3} (a). 

The case for the DQCP of JQ$_3$ ($\Delta=1.0$), $\ln X_M /L$ as a function of $\ln L$ seems to be consistent with that of a normal quantum critical point, while for the corner cut case it has been found that the sign of log coefficient $s$ in Eq.~\ref{eq:eq7} contradicts with unitary CFTs~\cite{wangScaling2022}. For the easy plane DQCPs ($\Delta<1.0$), one clearly sees that as $L$ is large enough there exist finite slopes and the slopes of $\ln |X_M(\theta)|/L$ became enhanced at the first-order phase transition points of EPJQ as anisotropy $\Delta \to 0$, as shown in the inset of Fig.~\ref{fig:fig3} (a). Therefore, the scaling behavior of disorder operator resemble that of residual N$\acute{e}$el orders, suggesting the first-order nature of the easy-plane DQCPs.

We have further computed a related quantity -- the bipartite spin fluctuation~\cite{songEntanglement2011}. At $\theta \to 0$ limit, the scaling of the disorder operator $-\ln |X_M|$ has a similar behavior to that of the bipartite spin fluctuations, which is defined as 
\begin{equation}
\label{eq:FM}
    F_{M}=\langle \left(S_M^z-\langle S_M^z\rangle\right)^2\rangle,~~~
    S_{M}^z=\sum_{i \in M} S_i^z.
\end{equation}
To be specific, at the  N$\acute{e}$el phase, the scaling of the bipartite fluctuations is $F_{M}\sim bL\ln L$~\cite{metlitskiEntanglement2011,songEntanglement2011}, and $F_{M}/L \sim b\ln L$ as show in Fig.~\ref{fig:fig3} (b). For the (2+1)D O(3) QCP of $J_1-J_2$ model, the bipartite spin fluctuations of the smooth region have a linear scaling $F_{M} \sim aL$ and $F_{M}/L$ will be a constant at large system size $L$. For the DQCP of JQ$_3$ model ($\Delta=1.0$), $F_{M}/L$ as a function of $\ln L$ gives a tiny slope at large system size $L$, suggesting a very tiny weakly first order nature of the transition point. More interestingly, the slopes of $F_M/L$ versus $\ln L$ also become larger at the phase transition points of the EPJQ as $\Delta$ decreases, especially at large system sizes and anisotropy $\Delta \to 0$ as highlighted by dashed rectangular boxes of Fig.~\ref{fig:fig3} (b). The scaling at large size reveals similar behavior as that of N$\acute{e}$el orders contributed from the remaining Goldenstone mode. This finding further strengthens that the DQCP transitions change from weak to prominent first-order transitions from $\Delta=1$ to $\Delta=0$.\\

\begin{figure}[htp!]
\centering
\includegraphics[width=\columnwidth]{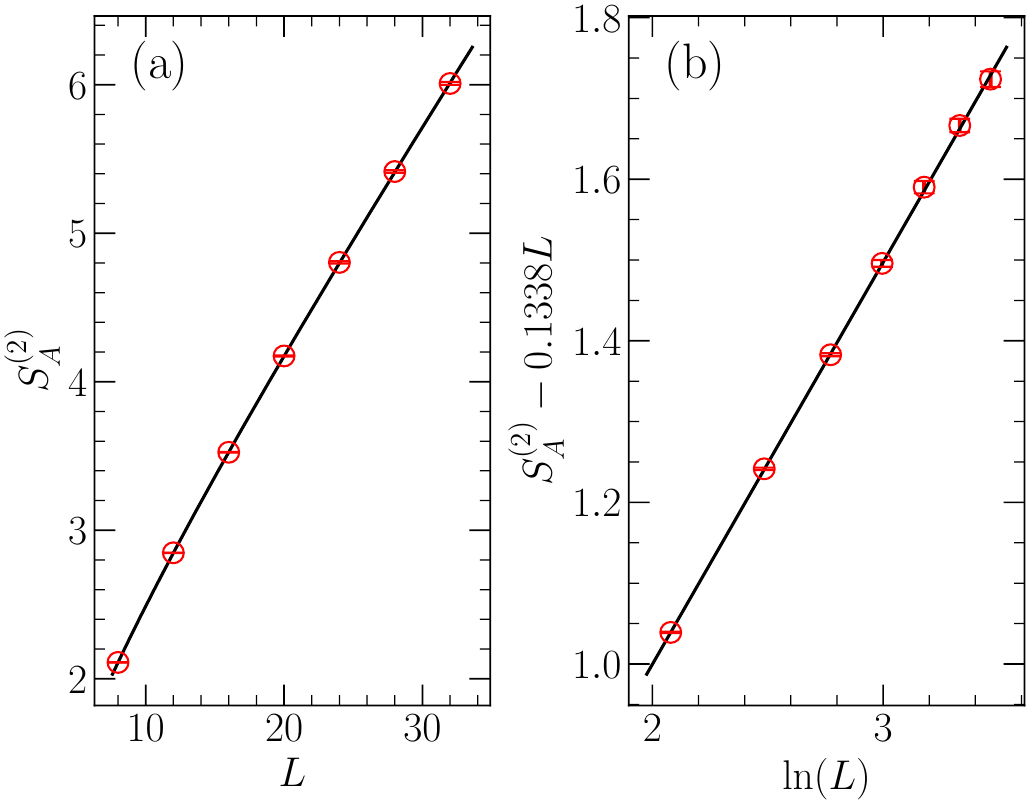}
\caption{Second R\'enyi entropy $S^{(2)}_{A}$ for $L\times L$ lattice with subsystem size of $L/2\times L$ deep in the AFXY phase of EPJQ model at $\Delta=0$ and $q=0.3$.(a) $S^{(2)}_{A}$ versus system size $L$. The black curve is a fit to $S^{(2)}_{A}=aL+b\ln(L)+c$ and the fitting result is $S^{(2)}_{A}=0.1338(8)L+0.497(11)\ln(L)+0.0046(166)$.  (b) $S^{(2)}_{A}-aL$ vs. $\ln(L)$ for different system sizes where we choose the fitted value $a=0.1338$. The black curve is a straight line of $y=0.497x+0.0046$.}
\label{fig:fig5-5}
\end{figure}

\begin{figure*}[htp!]
\centering
\includegraphics[width=2\columnwidth]{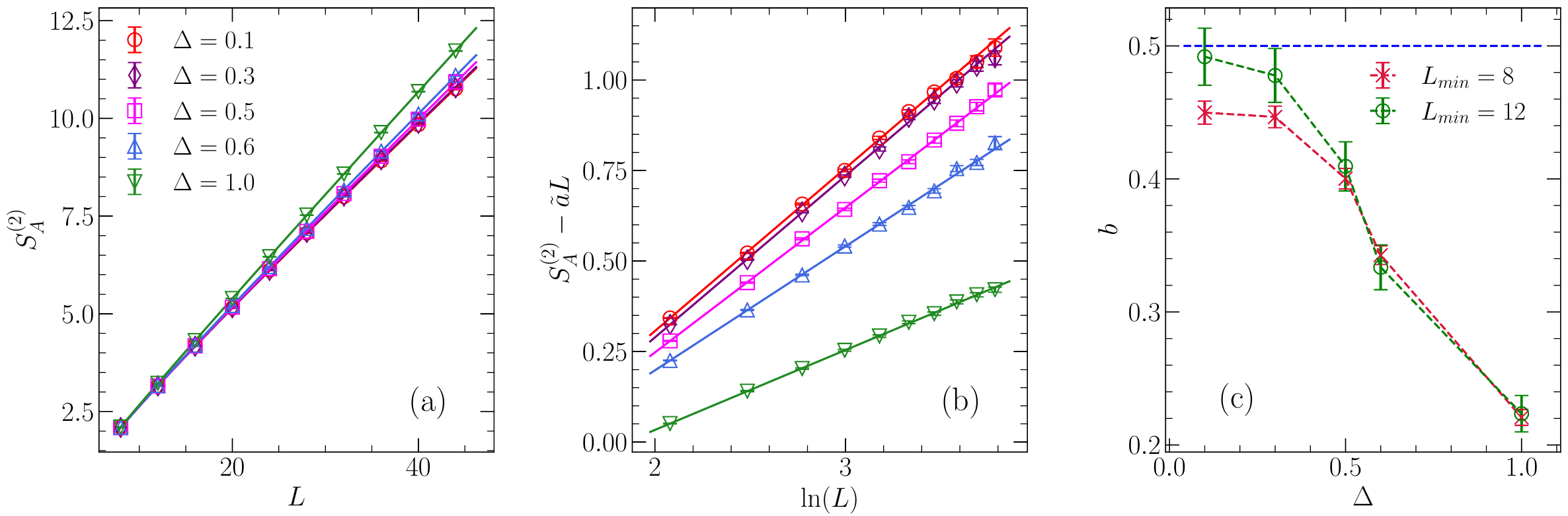}
\caption{Second R\'enyi entropy $S^{(2)}_{A}$ for $L\times L$ lattice with subsystem size of $L/2\times L$ at the phase transitions of EPJQ model for different anisotropy $\Delta$ as in Table.~\ref{qc} and $\Delta=1$ of JQ model. (a) $S^{(2)}_{A}$ versus $L$ for different $\Delta$. The solid lines are fitted curves of finite-size data according to Eq.~\eqref{eq:eq8}. (b) $S^{(2)}_{A}-\tilde{a}L$ versus $\ln(L)$ for different $\Delta$. $\tilde{a}=a(\Delta)$ is the fitted area law coefficient $a$ for different $\Delta$. (c) The fitted log-coefficient $b$ versus $\Delta$. We gradually exclude the data for small system sizes, and $L_{\text{min}}$ is the smallest system size used in the fitting process. The blue dotted line is a guide for eyes, indicating $b=0.5$ or $N_G =1$.}
\label{fig:fig6}
\end{figure*}

\section{Entanglement Entropy}
\noindent Next we investigate the EE at the transition points of various $\Delta$ of Eq.~\eqref{eq:eq1}. To this end, we employ the nonequilibrium increment method~\cite{albaOut2017,demidioEntanglement2020,zhaoScaling2022,zhaoMeasuring2022} within the framework of SSE-QMC simulation~\cite{Sandvik1999,Syljuaasen2002} to determine the second R\'enyi entropy of the EPJQ model at phase transition points for various values of $\Delta$. R\'enyi entanglement entropy is 
defined as $S_A^{(n)}=\frac{1}{1-n}\ln \Tr \rho_A^{n}$ which can be reexpressed in the form of $S_A^{(n)}=\frac{1}{1-n}\ln \frac{Z_{A}^{(n)}}{Z^{(n)}}$ according to the replica trick~\cite{calabreseEntanglement2004}. 
The nonequilibrium method is based on Jarzynski's equality~\cite{Jarzynski1997}, which relates the free energy difference between two systems with the total work done during a tuning process from one system to another. We regard the partition functions $Z_{A}^{(n)}$ and $Z^{(n)}$ as those of two different physical systems, then it is natural to apply the Jarzynski's equality and design a tunning process between the two systems to calculate the R\'enyi entropies~\cite{albaOut2017,demidioEntanglement2020}. In practice, we follow the incremental version~\cite{zhaoScaling2022,zhaoMeasuring2022} of Ref.~\cite{demidioEntanglement2020},  and it can overcome the obstacles that the EE is in general an exponential observable~\cite{zhangIntegral2024,zhouIncremental2024}. We conduct the simulation  on a $L\times L$ square lattice with periodic boundary conditions, and fix with $\beta=L$ and choose the subregion A to be a $L\times L/2$ cylinder defined same as region $M$ in the right inset of Fig.~\ref{fig:fig1} (a), which has no sharp corners on the entanglement boundary. 

For $\Delta<1.0$, in the AFXY phase, the system spontaneously breaks the SO(2) spin symmetry and possesses one Goldstone mode. In this case, EE is expected to scale as~\cite{metlitskiEntanglement2011}
\begin{equation}
\label{eq:eq8}
    S^{(n)}_{A}=aL+b\ln(L)+c, \quad b=\frac{N_g}{2},
\end{equation}
where $N_g$ represents the number of Goldstone modes. Eq.~\eqref{eq:eq8} has been extensively verified for both XY phase with one Goldstone mode and N\'eel ordered states with two Goldstone modes respectively~\cite{Helmes2014,kulchytskyyDetecting2015,demidioEntanglement2020,zhaoMeasuring2022,songExtracting2023,Deng2023improved,zhouIncremental2024,D'Emidio2024universal} . We first test our algorithm in the AFXY phase of the EPJQ model at $\Delta=0$ and $q=0.3$ deep in the AFMXY phase, with the results displayed in Fig.~\ref{fig:fig5-5}. We measure the second order R\'enyi EE for the  $L\times L/2$ subsystem $A$ on a $L\times L$ square lattice with smooth boundary. The system sizes utilized are $L=8,12,16,\ldots,32$ and the temperature is fixed at $\beta=L$. As shown in Fig.~\ref{fig:fig5-5}(b), by fitting the finite-size EE data to Eq.~\eqref{eq:eq8}, we observe a good agreement between our data with the scaling form of Eq.~\eqref{eq:eq8}. Moreover, the log-coefficient obtained from curve fitting is $\frac{N_G}{2}=0.497\pm0.011$, aligning well with the expectation of existence of one Goldstone mode, which corresponds to spontaneous SO(2) symmetry breaking. The analysis of the fitting quality can be found in the Appendix~\ref{sec:SMII}.

 We then proceed to the phase transition points.  Our extrapolated finite order parameter data in Fig.~\ref{fig:fig1}(b)  shows that the first-order behavior is more pronounced as $\Delta$ decreases from 1 to 0. However, the relation between the scaling of entanglement entropy (EE) with the strengthen of first-order behavior remains unexplored. We measure the second R\'enyi EE at the transition points in Table.~\ref{qc} for various values of $\Delta$ and system sizes $L=8,12,16,\ldots,44$ and fix $\beta=L$. The subsystem A is again chosen to be $L\times L/2$ cylinder withour corners on the boundary.
 As presented in Fig.~\ref{fig:fig6}, for all examined values of $\Delta$, we observe a clear logarithmic correction to the area law term in the scaling of EE (Eq.~\eqref{eq:eq8}). Additionally, as $\Delta$ decreases and the first-order behavior strengthens, the fitted log-coefficient $b$ increases. Interestingly, for $\Delta=0.1$ and $\Delta=0.3$, the fitted value of $b$ is even consistent with 0.5, which corresponds to the existence of a single Goldstone mode, the same as the scaling of EE deep inside the AFXY phase in Fig.~\ref{fig:fig5-5}. Our findings suggest that, for $\Delta\le 0.3$, the logarithmic correction from single Goldstone mode can also be observed even in the mixed phase of N\'eel order and VBS state. For $\Delta > 0.3$ and as $\Delta$ increases, the first-order behavior becomes weaker and weaker as the remaining order parameters suggest, it is very likely here the fitted $b$ suffers from strong finite-size effects and will converge to 0.5 as system size grows. Here with the limited system sizes, it is impossible to address this issue thoroughly. We leave this issue for further studies.\\

\section{Discussion}
\noindent Our work shows the two non-local measurements -- the disorder operator and entanglement entropy -- consistently exhibit anomalous scaling behaviors at the easy-plane deconfined quantum criticalities. By adjusting the anisotropy from the Heisenberg limit ($\Delta=1$) to the easy-plane limit ($\Delta=0$), we find that as the first-order nature of the transition is amplified, the anomalous behavior for both  disorder operator and entanglement entropy are also strengthened, and they resembles contributions from Goldstone modes. Interestingly, when $\Delta\le 0.3$, the log coefficients in EE gradually converge to $b=0.5$ which is consistent with one Goldstone mode and SO(2) symmetry breaking at the first-order transition points. Our work thus provides strong evidence that the observed anomalous scaling behaviors of the entanglement measurements (disorder operator and EE) indeed come from the weakly-first-order nature of DQCPs of JQ model realizations both at $\Delta=1$ and $\Delta<1$.

Future research directions may include investigating the disorder operator and entanglement entropy scaling of JQ$_n$ models with $n>3$, which have been shown to exhibit stronger first-order transitions~\cite{Takahashi2020,takahashiSO52024} as $n$ increases, similar to EPJQ models. Another avenue of interest is to explore the entanglement entropy scaling of an absolutely strong first-order transition, which involves a mixture of two phases with classical probabilities. Moreover, multipartite entanglement~\cite{osborneEntanglement2002,javanmardSharp2018,wangEntanglement2024,parezFate2024} and higher-order R\'enyi entropies, entanglement negativity and entanglement spectrum~\cite{wuEntanglement2020,wangEntanglement2023,yanUnlocking2023} with QMC simulations at the DQCP also present intriguing topics for future investigation.

\section*{Acknowledgement}
We thank Meng Cheng, Cenke Xu, Anders Sandvik, Senthil Todadri, Subir Sachdev for valuable discussions on the related topic. NVM and ZYM acknowledge the earlier insightful discussions on the transitions of EPJQ model with Arnab Sen and Anders Sandvik. JRZ and ZYM acknowledge the support from the Research Grants Council (RGC) of
Hong Kong Special Administrative Region of China (Project Nos. 17301721, AoE/P-701/20, 17309822, HKU C7037-
22GF, 17302223), the ANR/RGC Joint Research Scheme sponsored by RGC of Hong Kong and French National Research Agency (Project No. A HKU703/22), the GD-NSF (No. 2022A1515011007) and the HKU Seed Funding for
Strategic Interdisciplinary Research. Y.C.W. acknowledges the support from the Natural Science Foundation of China (Grant No. 12474216), the Zhejiang Provincial Natural Science Foundation of China (Grant No. LZ23A040003), and the start-up funding and the High-Performance Computing Centre of Hangzhou International Innovation Institute of Beihang University. NVM acknowledges the National Natural Science Foundation of China (No.
12004020) and the Fundamental Research
Funds for the Central Universities.
We thank HPC2021 system under the Information Technology Services and the Blackbody HPC system at the Department of Physics, University of Hong Kong, as well as the Beijng PARATERA Tech CO.,Ltd. (URL: https://cloud.paratera.com) for providing HPC resources that have contributed to the research results reported within this paper.

\bibliography{epjq}

\clearpage
\onecolumngrid
\appendix



\section{The criticality for different $\Delta$ in the EPJQ models}
\label{sec:SMI}

\begin{figure*}[htp!]
\centering
\includegraphics[width=\columnwidth]{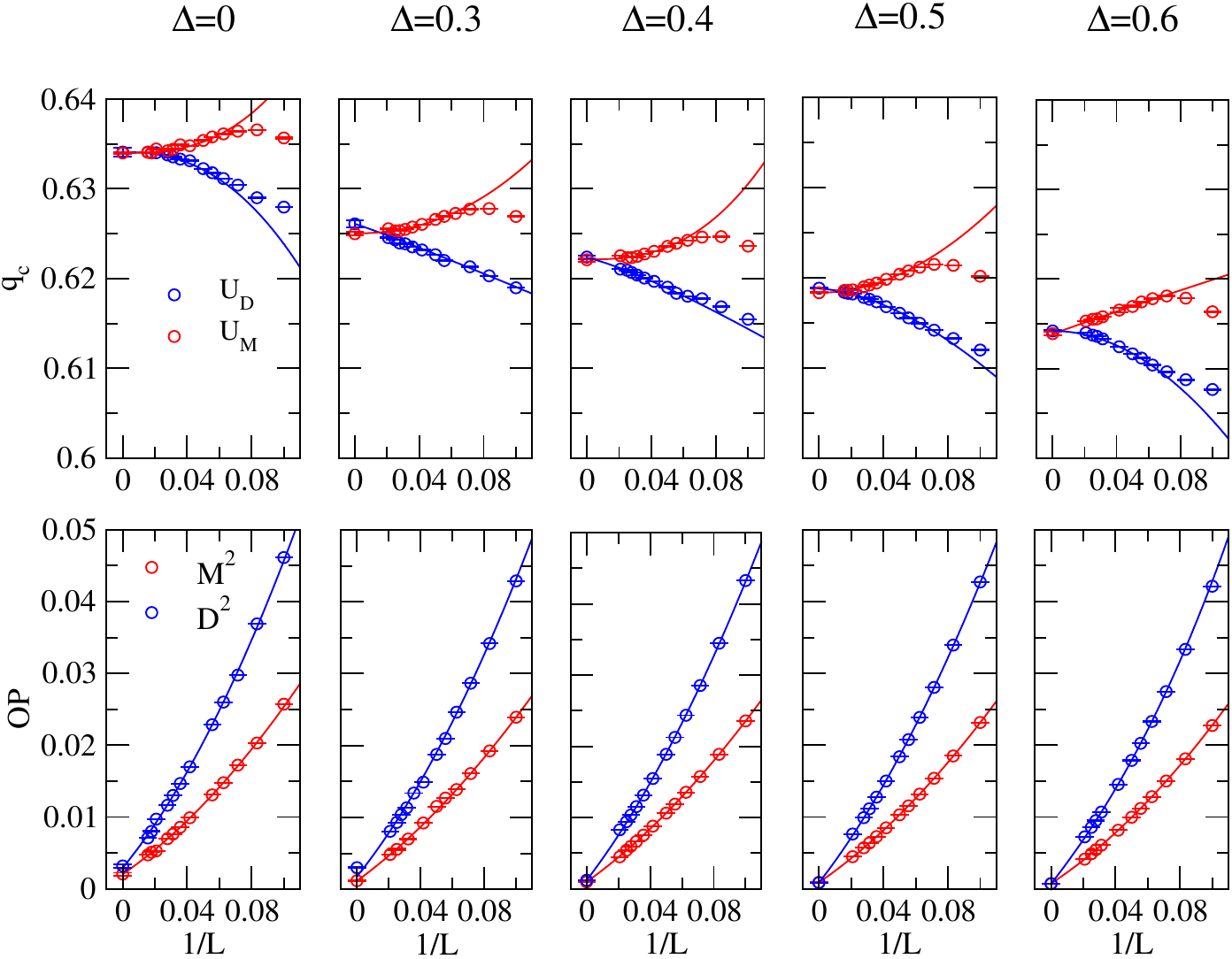}
\caption{(Upper panels) The crossing points $q_{c}(L)$ obtained from the curve of dimensionless Binder cumulants $U_{M}$(blue) and $U_{D}$(red) versus $q$ close to the phase transition points for two system sizes $L$ and $2L$. The red/blue curve is the fitting function with the scaling form of $q_{c}(L)$ in Eq.~\eqref{eq:eq5} excluding the crossings for small sizes with $\chi^{2}$ of the fitting becomes close to one. The data points presented at $1/L=0$ are the y-intecepts of fitting curves with the fitting errors which are given in Tab.\ref{qc}. (Lower panels) The value of order parameters for both AFXY and VBS phases at $q_c(L)$ obtained with corresponding Binder cumulants versus $1/L$ in the EPJQ models with different $\Delta$ for all computing sizes $L$. The red/blue curve is the second-order polynormial fitting function of all the data points with $\chi^{2}$ of the fitting close to one. The data points presented at $1/L=0$ are the y-intecepts of the fitting curves with errors illustrated in Fig.~\ref{fig:fig1}.}
\label{fig:difflam}
\end{figure*}

The results in (b) of Fig.~\ref{fig:fig1} in the main text present that the order parameters for both  the AFXY and VBS phases converge to a non-zero value when $L$ goes to infinity for different $\Delta$ in the EPJQ models, which implies the possibility of first-order phase transitions in the EPJQ model. In this section we present the detail procedure in obtaining those converged order parameters at thermodynamic limit in Fig.~\ref{fig:fig1}(b). 
The binder cumulants $U_{M}$ and $U_{D}$ defined in Eq.~\eqref{eq:eq4} for the  AFXY and VBS phases correspondingly are two dimensionless quantities that are commonly chosen in the FSS of calculating critical points and correlation length exponents. In this paper we study the critical behavior of the EPJQ model with different anisotropic value $\Delta$
with the help of $U_{M}$ and $U_{D}$ all the crossing points $q_{c}(L)$ of two simulated sizes $L$ and $2L$ got from $U_{M}$ and $U_{D}$  are shown in the first row in Fig.~\ref{fig:difflam} for those different $
\Delta$. Using the fitting form in Eq.~\eqref{eq:eq5} two $q_{c}(\infty)$ can be got, which are listed in Tab.\ref{qc} in the main text. At each crossing point $q_{c}(L)$ we also calculate the square of order parameters defined in Eq.~\ref{eq:eq2} and Eq.~\ref{opall} as $M_{x}^{2}(L)$ and $D^{2}(L)$ illustated in the second row in Fig.~\ref{fig:difflam}. It should be noticed that $M_{x}^{2}(L)$ and $D^{2}(L)$ are calculated at different  $q_{c}(L)$ as the crossing points are got from different dimensionless quantities. In the calculation of $M_{x}^{2}(L)$ we chose the corresponding $q_{c}(L)$ got from $U_{M}$ for two sizes $L$ and $2L$. As for the $D^{2}(L)$ the crossing points are located using $U_{D}$. After all the $M_{x}^{2}(L)$ and $D^{2}(L)$ are known for different sizes, the square of order parameters at critical points when $L\rightarrow \infty$ can be calculated with the second-order polynomial fitting. All the fitting parameters of $M_{x}^{2}(\infty)$ and $D^{2}(\infty)$ for different $\Delta$ are illustrated in Fig.~\ref{fig:fig1} in the main text.

\section{Comparison of fitting quality}
\label{sec:SMII}
In the main text, the scaling form of EE we use for the fitting is 
\begin{equation}
\label{fitting-1}
    S^{(2)}_{A}=aL+b\ln(L)+c.
\end{equation}
However, to confirm the validity of the fitting form we use, we consider the comparison of above fitting form with the following one
\begin{equation}
\label{fitting-2}
    S^{(2)}_{A}=aL+b/L+c,
\end{equation}
which has no subleading $L-$ dependence term but with a finite-size correction term $b/L$.\begin{figure}[htp!]
\centering
\includegraphics[width=0.55\columnwidth]{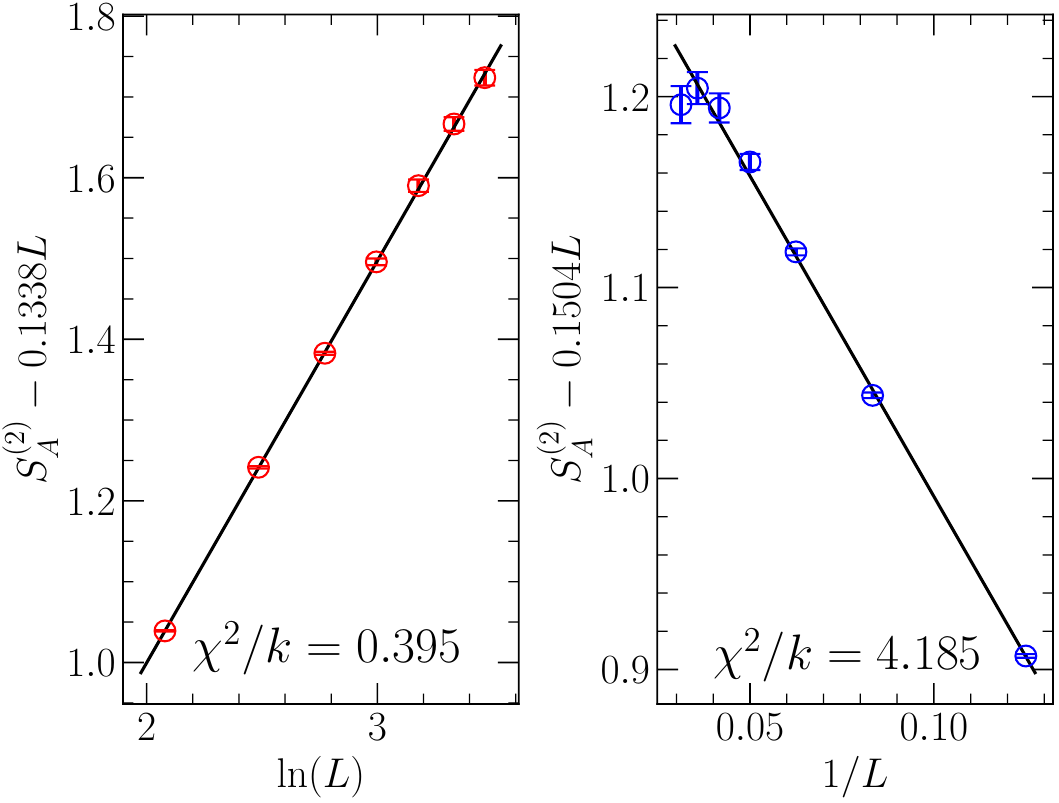}
\caption{Comparison of fitting quality with Eq.~\ref{fitting-1} and Eq.~\ref{fitting-2} in the AFMXY phase at $\Delta=0$ and $q=0.3$. Left panel: $S^{(2)}_{A}- 0.1338L$ versus $\ln(L)$ with area law coefficient obtained from  fitting to Eq.~\ref{fitting-1}. Right panel: $S^{(2)}_{A}- 0.1504L$ versus $1/L$ with area law coefficient obtained from  fitting to Eq.~\ref{fitting-2}. The chi-squared values of two fittings are  also printed on the figure. }
\label{fig:figs3}
\end{figure}
\begin{figure*}[htp!]
\centering
\includegraphics[width=\columnwidth]{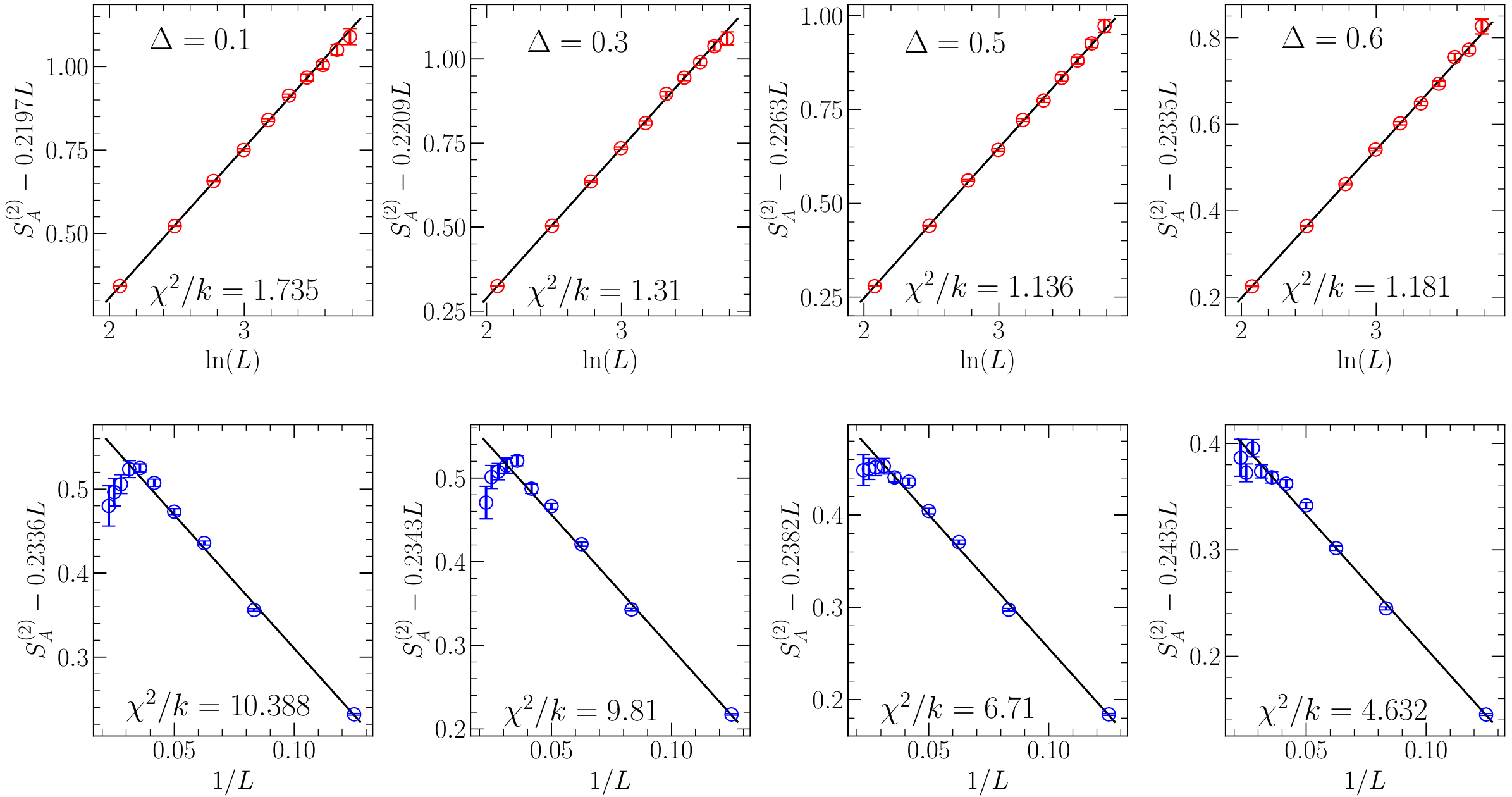}
\caption{Comparison of fitting quality with Eq.~\ref{fitting-1} and Eq.~\ref{fitting-2} at the DQCPs of easy-plane $JQ_3$ model. The red and blue data points represent the finite-size EE data with an area law term subtracted. The area law term is fitted from Eq.~\ref{fitting-1} for red data points and Eq.~\ref{fitting-2} for blue data points.}
\label{fig:figs4}
\end{figure*}

We use the chi-squared value $\chi^{2}/k$ to evaluate the fitting quality. chi-squared value is defined as
\begin{equation}
        \chi^2=\sum_{i=1}^N \frac{\left(f(x_{i})-y_{i}\right)^2}{\sigma_i^2 }, 
\end{equation}
where $\sigma_i$ is the uncertainty of data $y_i$. $k = N - r$ is the effective number of degrees of freedom, where $N$ is the number of data points and $r$ is the number of fitting parameters. $\chi^2/k$ is expected to be close to 1 for a good fit. As shown in Fig.~\ref{fig:figs3}, in the AFMXY phase, the fitting quality is compared with Eq.~\ref{fitting-1} and Eq.~\ref{fitting-2} and the $\chi^{2}/k$ values are 0.395 and 4.185 separately. $\chi^2/k$ should be typically distributed within the range $[1-\sqrt{2/k},1+\sqrt{2/k}]$ where for this case the range is  $[0.465,1.535]$. In our simulation, the precise estimation of statistic errors can be affected by limited number of Monte Carlo bins so that we attribute the small deviation of 0.395 to 0.465 to problematic errorbars and 4.185 is apparently far away from the ideal range.  We thus conclude that Eq.~\ref{fitting-1} fits the data better than Eq.~\ref{fitting-2}.

Similarly, for DQCPs of easy-plane $JQ_3$ model, the comparison of fitting quality with Eq.~\ref{fitting-1} and Eq.~\ref{fitting-2} is also performed and the results are shown in Fig.~\ref{fig:figs4}. By comparing their $\chi^{2}/k$ values, we conclude that Eq.~\ref{fitting-1} in all cases fit better than Eq.~\ref{fitting-2}. Note that for $\Delta=1.0$ which is the standard $JQ_3$ model, it has already been shown in previous study~\cite{songExtracting2023} that Eq.~\ref{fitting-1} fits the better. In that case we only present the comparison for $\Delta<1.0$ in this manuscript.

\end{document}